\documentclass[epj]{svjour}
\usepackage{graphics}
\usepackage{amssymb}
\usepackage{amsmath}
\usepackage[dvips]{epsfig}
\newcommand{\fmn}[2]{\mbox{${\textstyle \frac{#1}{#2}}$}}
\newcommand{\dd}{\mbox{\rm d}}

\newcommand{\half}{\mbox{${\textstyle \frac{1}{2}}$}}

\authorrunning{Fran\c{c}ois Lehar, Colin Wilkin}

\begin{document}
\title{Nucleon charge exchange on the deuteron: A critical review}
\author{Fran\c{c}ois Lehar\inst{1}$^,$\inst{2}
       \and
       Colin Wilkin\inst{3}}
\institute{SPP IRFU, CEA Saclay, F-91190 Gif-sur-Yvette Cedex, France
       \and IEAP CTU, Horsk\'a 3a/22, Cz-12800 Prague 2, Czech Republic,
            \email{lehar@mail.utef.cvut.cz}
       \and Physics and Astronomy Department, UCL, Gower Street, London WC1E 6BT, United Kingdom,
            \email{cw@hep.ucl.ac.uk}}
\date{Received: \today}
\abstract{The existing experimental data on the $d(n,p)nn$ and
$d(p,n)pp$ cross sections in the forward direction are reviewed in
terms of the Dean sum rule. It is shown that the measurement of
the ratio of the charge exchange on the deuteron to that on the
proton might, if taken together with other experimental data,
allow a direct construction of the $np\to np$ scattering amplitude
in the backward direction with few ambiguities.
} 
%
%
\maketitle
\section{Introduction}
\label{intro} \setcounter{equation}{0}

The nucleon-nucleon interaction is fundamental to the whole of
nuclear physics and hence to the composition of matter as we know
it. Apart from its intrinsic importance, it is also a necessary
ingredient in the description of meson production and other
intermediate energy processes.

In the case of proton-proton scattering, the data set of
differential and total cross sections and the various single and
multi-spin observables is very extensive and this has allowed the
construction of reliable isospin $I=1$ phase shifts up to at least
2\,GeV~\cite{SAID}. The situation is far less developed for the
isoscalar $I=0$ case, where the corresponding phase shift analysis
is only available up to 1.3\,GeV and even then there are
significant ambiguities at the higher energies~\cite{SAID}.

More good data on neutron-proton scattering are clearly needed,
possibly with the aim of directly reconstructing the isosinglet
amplitudes. This is particulary promising in the forward
direction~\cite{Ball1} but the conditions are almost as favourable
for backward $pn\to np$ scattering (often loosely called the $np$
charge-exchange region) and it is this which we want to consider
in some detail in this paper.

To avoid some of the problems associated with the quality of
neutron beams and/or the detection of neutron, the deuteron is
often used successfully as a substitute target or beam. For
example, it has been shown that the spin correlation and transfer
parameters in $pp$ quasi-elastic scattering in the 1.1 to 2.4~GeV
range are very close to those measured in free $pp$
collisions~\cite{Ball2} and the Saclay group find exactly the same
reassurance for $pn$ quasi-elastic scattering~\cite{Lesquen}.

One particular valuable tool that can be used to study the
backward amplitudes is the comparison of the quasi-free $(p,n)$ or
$(n,p)$ reaction on the deuteron to the free backward elastic
scattering on a nucleon target. It was emphasised over 50 years
ago that the reaction on the deuteron can act, in suitable
kinematic regions, as a spin filter that selects the
spin-dependent contribution to the $np$ elastic cross
section~\cite{Pomeranchuk}. This sensitivity arises from the Pauli
principle, which blocks any spin-1 component in the low energy
$\{nn\}$ or $\{pp\}$ system. To avoid the explicit introduction of
the dynamics of the low energy $NN$ system, Dean~\cite{Dean}
derived a sum rule for the ratio $R_{np}$ of the differential
cross section for charge exchange on the deuteron to that on the
nucleon. Although this is given as a function of the momentum
transfer $q$, it simplifies for collinear dynamics to the extent
that there is then no dependence on the deuteron structure. More
importantly, the sum rule converges very fast as a function of the
excitation energy for small $q$, due to the strength of the low
energy $^{1\!}S_0$ $NN$ interaction. It has therefore been used in
the analysis of the wealth of $R_{np}$ data, which now extend up
to 2\,GeV~\cite{Sharov08}. It is the aim of the present paper to
show how such data, combined with other measurements, might
contribute to an analysis of the elastic neutron-proton scattering
amplitudes in the backward direction.

In section~\ref{Amps} we summarise the amplitudes and some
observables that are relevant for backward elastic $np$
scattering. Of particular importance in this context is the fact
that the conventional $NN$ amplitudes~\footnote{See
Ref.~\cite{BYS78} for a very comprehensive discussion of different
amplitude bases.} are not the most suitable ones when analysing
charge exchange on the deuteron, where it is necessary to take
into account of the interchange between the final neutron and
proton \textit{ab initio}. The impulse approximation dynamics and
the form of the Dean sum rule in the forward direction are
described in section~\ref{deuteron}, where some of the underlying
assumptions are clarified. It is shown there that the sum rule
saturates very quickly, which make it such a useful tool.

Section~\ref{data} gives an extensive compilation of the values of
$R_{np}(0)$ derived from the $nd \to p\{nn\}$, $pd\to n\{pp\}$,
and $dp\to \{pp\} n$ reactions and it is shown there that the
total error bars are the smallest in the $(n,p)$ case provided
that a good quality neutron beam is available. On the basis of the
existing phase shift analysis~\cite{SAID}, impulse approximation
predictions of $R_{np}(0)$ can be made up to a beam energy of
1.3\,GeV and the agreement with experimental data is very
reasonable down to at least 300\,MeV. However, it is important to
reiterate that values of $R_{np}(0)$ are now available up to
2\,GeV. The prospects for a $np \to np$ elastic amplitude
reconstruction in the backward direction are discussed in the
conclusions of section~\ref{summary}, where it is shown that, with
extra information available through the use of polarised
deuterons, this is now becoming feasible.

%
%
\section{Neutron-proton amplitudes and observables}
\label{Amps} \setcounter{equation}{0}

The nucleon-nucleon formalism, including the four-index notation
and definition of all \textit{pure} experiments, is discussed in
full detail in Ref.~\cite{BYS78}. In this work the matrix
describing
elastic neutron-proton scattering is written in the form%
\begin{eqnarray}%
\nonumber M(\vec{k}_f,\vec{k}_i) &=& \half\big[(a + b) + (a -
b)(\vec{\sigma}_n\cdot\hat{\vec{n}})(\vec{\sigma}_p\cdot\hat{\vec{n}})\\
\nonumber
&&\hspace{-2cm}+(c+d)(\vec{\sigma}_n\cdot\hat{\vec{m}})
(\vec{\sigma}_p\cdot\hat{\vec{m}})
+(c-d)(\vec{\sigma}_n\cdot\hat{\vec{\ell}})(\vec{\sigma}_p\cdot
\hat{\vec{\ell}})\\ &&+e(\vec{\sigma}_n + \vec{\sigma}_p)\cdot\hat{\vec{n}}\big]\,,%
\label{BYS}
\end{eqnarray}%
where $a$, $b$, $c$, $d$ and $e$ are complex invariant amplitudes,
which are functions of energy and scattering angle $\theta$. The
$2\times 2$ Pauli matrices $\vec{\sigma}_n$ and $\vec{\sigma}_p$
act in the spaces of the proton and neutron spins, respectively.

In terms of the c.m.\ momenta in the initial and final states,
$\vec{k}_i$ and $\vec{k}_f$, an orthonormal basis system is
defined through%
\begin{equation}\label{basis1}%
\hat{\vec{n}} =
\frac{\vec{k}_i\times\vec{k}_f}{|\vec{k}_i\times\vec{k}_f|}\,,~~~~
 \hat{\vec{\ell}} =
\frac{\vec{k}_f + \vec{k}_i}{|\vec{k}_f+\vec{k}_i|}\,,~~~~
 \hat{\vec{m}} =
\frac{\vec{k}_f - \vec{k}_i}{|\vec{k}_f-\vec{k}_i|}\,,
\end{equation}
which satisfy $\hat{\vec{m}}=\hat{\vec{n}}\times\hat{\vec{\ell}}$.

For later purposes, it is convenient to choose the invariant
normalisation, where the unpolarised cross section is given by%
\begin{equation}
\left(\frac{\dd\sigma}{\dd t}\right)_{\!np\to np} =
\half\left(|a|^2+|b|^2+|c|^2+|d|^2+|e|^2\right),
\end{equation}
where $t$ is the four-momentum transfer between the initial and
final neutrons.

In the forward direction $e=0$ and, since one can then not
distinguish between the two perpendicular axes,
$a(0)-b(0)=c(0)+d(0)$. The scattering matrix then reduces to %
\begin{eqnarray}%
\nonumber M(\vec{k}_i,\vec{k}_i) &=& \half\big[(a(0) + b(0)) +
(c(0)-d(0))(\vec{\sigma}_n\cdot\hat{\vec{\ell}})(\vec{\sigma}_p\cdot
\hat{\vec{\ell}})\\ \nonumber &&\hspace{-1.5cm} +(a(0) -b(0))
\left\{(\vec{\sigma}_n\cdot\hat{\vec{n}})(\vec{\sigma}_p\cdot\hat{\vec{n}})
+(\vec{\sigma}_n\cdot\hat{\vec{m}})(\vec{\sigma}_p\cdot\hat{\vec{m}})\right\}
\big]\\ %
\end{eqnarray}%

There are three spin-correlated total cross sections defined by%
\begin{equation}
\sigma_{\rm tot}=\sigma_0 -\half\Delta\sigma_{L}\, \mathcal{P}_n^L
\mathcal{P}_p^L - \half\Delta\sigma_T\, \vec{\mathcal{P}}_n^T\cdot
\vec{\mathcal{P}}_p^T\,,
\end{equation}
where $\mathcal{P}^L$ and $\vec{\mathcal{P}}^T$ are the
longitudinal and transverse components of the polarisation of
either the initial neutron or proton.

The imaginary parts of the three independent forward amplitudes
can be determined through measurements of these total cross
sections using the relations:
\begin{eqnarray}
\nonumber\sigma_0 &=& 2\sqrt{\pi}\,\textit{Im}[a(0)+b(0)]\,,\\
\nonumber -\Delta\sigma_{T} &=& 4\sqrt{\pi}\,\textit{Im}[a(0)-b(0)]\,,\\
-\Delta\sigma_L &=& 4\sqrt{\pi}\,\textit{Im}[c(0)-d(0)]\,.
\end{eqnarray}

We are interested in backward rather than forward neutron-proton
scattering, in a region that is often called neutron-proton charge
exchange. The interchange of the momenta of the final neutron and
proton is achieved by letting $\vec{k}_f \to -\vec{k}_f$ and, as
is seen from Eq.(\ref{basis1}), this introduces a new set of basis
vectors, which are related to the original ones through
\begin{equation}\label{basis2}
\hat{\vec{n}}_{ce} = - \hat{\vec{n}}\,,~~~\hat{\vec{\ell}}_{ce} =
- \hat{\vec{m}}\,,~~~\hat{\vec{m}}_{ce} = - \hat{\vec{\ell}}\,.
\end{equation}

The scattering matrix in this representation becomes
\begin{eqnarray}%
\nonumber M(-\vec{k}_f,\vec{k}_i) &=& \half\big[(a + b) + (a -
b)(\vec{\sigma}_n\cdot\hat{\vec{n}}_{ce})(\vec{\sigma}_p\cdot\hat{\vec{n}}_{ce})\\
\nonumber
&&\hspace{-2cm}+(c+d)(\vec{\sigma}_n\cdot\hat{\vec{\ell}}_{ce})
(\vec{\sigma}_p\cdot\hat{\vec{\ell}}_{ce})
+(c-d)(\vec{\sigma}_n\cdot\hat{\vec{m}}_{ce})(\vec{\sigma}_p\cdot
\hat{\vec{m}}_{ce})\\
&&-e(\vec{\sigma}_n + \vec{\sigma}_p)\cdot\hat{\vec{n}}_{ce}\big]\,,%
\end{eqnarray}%
which, in the strictly backward direction of $\theta=\pi$, reduces
to
\begin{eqnarray}%
\nonumber M(-\vec{k}_i,\vec{k}_i) &=& \half\big[(a(\pi) +
b(\pi))+\\
\nonumber &&\hspace{-2.3cm}
(c(\pi)+d(\pi))(\vec{\sigma}_n\cdot\hat{\vec{\ell}}_{ce})(\vec{\sigma}_p\cdot
\hat{\vec{\ell}}_{ce})+\\ &&\hspace{-2.3cm}(a(\pi) -b(\pi))
\left\{(\vec{\sigma}_n\cdot\hat{\vec{n}}_{ce})(\vec{\sigma}_p\cdot\hat{\vec{n}}_{ce})\right.\nonumber\\
&&\hspace{-0.5cm}\left.+(\vec{\sigma}_n\cdot\hat{\vec{m}}_{ce})(\vec{\sigma}_p\cdot\hat{\vec{m}}_{ce})\right\}
\big]\,, %
\end{eqnarray}%
where $e(\pi)=0$ and $a(\pi)-b(\pi)=c(\pi)-d(\pi)$.

If one invokes the symmetry properties of the amplitudes that
follow from isospin invariance~\cite{BYS78}, the backward values
of the imaginary parts of the three independent amplitudes are, in
principle, determined by the values of the spin-dependent total
cross sections for neutron-proton and proton-proton scattering:
\begin{eqnarray}
\nonumber
\textit{Im}[a(\pi)]&=&\phantom{-}\frac{1}{8\sqrt{\pi}}\left(2\sigma_0^{(-)}
-\Delta\sigma_T^{(-)}\right),\\
\nonumber
\textit{Im}[b(\pi)]&=&\phantom{-}\frac{1}{8\sqrt{\pi}}\left(\Delta\sigma_T^{(-)}
+\Delta\sigma_L^{(-)}\right),\\
\nonumber
\textit{Im}[c(\pi)]&=&\phantom{-}\frac{1}{8\sqrt{\pi}}\left(2\sigma_0^{(-)}
+\Delta\sigma_T^{(-)}\right),\\
\textit{Im}[d(\pi)]&=&-\frac{1}{8\sqrt{\pi}}\left(\Delta\sigma_T^{(-)}
-\Delta\sigma_L^{(-)}\right),
\end{eqnarray}
where we use the notation
\begin{equation}
\sigma^{(-)} \equiv \sigma(np)-\sigma(pp)
\end{equation}
for all three total cross sections.

However, in the theoretical treatment of the charge exchange on
the deuteron, \textit{i.e.}\ $nd\to p\{nn\}$ at small angles
between the incident neutron and final proton, it is convenient to
work with an alternative amplitude decomposition~\cite{BW}:
\begin{eqnarray}
\nonumber
M^{ce}(\vec{k}_f,\vec{k}_i)&=&\mathcal{P}_{p\leftrightarrow
n}\big[\alpha^{ce} +i\gamma^{ce}
(\vec{\sigma}_{n}+\vec{\sigma}_{p})\cdot\hat{\vec{n}}_{ce}\\
\nonumber &&\hspace{-2.5cm}+\beta^{ce} (\vec{\sigma}_{n} \cdot
\hat{\vec{n}}_{ce})(\vec{\sigma}_{p} \cdot\hat{\vec{n}}_{ce})+
\delta^{ce} (\vec{\sigma}_{n} \cdot
\hat{\vec{m}}_{ce})(\vec{\sigma}_{p} \cdot \hat{\vec{m}}_{ce})\\
\label{Mce} &&\hspace{-2.5cm}+\epsilon (\vec{\sigma}_{n} \cdot
\hat{\vec{\ell}}_{ce})(\vec{\sigma}_{p} \cdot
\hat{\vec{\ell}}_{ce})\big],
\end{eqnarray}
where the operator $\mathcal{P}_{p\leftrightarrow n}$ interchanges
the charge labels on the final proton and neutron. The presence of
this operator means that $\alpha^{ce}$ represents the
spin-independent amplitude between the initial neutron and final
proton whereas the $(a+b)$ of Eq.~(\ref{BYS}) corresponds to the
spin-independent amplitude between the initial and final neutrons.

It is straightforward to find the relationship between these two
representations and, for the collinear situation that is of
interest to us here, we have
\begin{eqnarray}
\nonumber \alpha^{ce}(0)&=&\fmn{1}{2}\left(a(\pi)+c(\pi)\right)\,,\\
\nonumber \beta^{ce}(0)=\delta^{ce}(0)&=&\fmn{1}{2}\left(a(\pi)-c(\pi)\right)\,,\\
\epsilon^{ce}(0)&=&\fmn{1}{2}\left(b(\pi)+d(\pi)\right)\,.
\label{identification2}
\end{eqnarray}

The relation between the imaginary parts of these forward
amplitudes and the total cross sections is more intuitively
obvious:
\begin{eqnarray}
\nonumber
\textit{Im}[\alpha^{ce}(0)]&=&\phantom{-}\frac{1}{4\sqrt{\pi}}\,\sigma_0^{(-)},\\
\nonumber
\textit{Im}[\beta^{ce}(0))]&=&-\frac{1}{8\sqrt{\pi}}\,\Delta\sigma_T^{(-)},\\
\label{sigtot2}
\textit{Im}[\epsilon^{ce}(0)]&=&\phantom{-}\frac{1}{8\sqrt{\pi}}\,
\Delta\sigma_L^{(-)}.
\end{eqnarray}

Extra information on the real parts of these amplitudes might be
obtained through the use of forward dispersion relations. This is
likely to be of most value for the spin-independent term, which
only involves unpolarised total cross section input. However, this
approach will not be pursued here.

From the relations given in Ref.~\cite{BYS78}, the magnitudes of
these charge-exchange amplitudes in the forward direction are
given in terms of the backward elastic $np$ differential cross
section and spin-transfer parameters $K_{0nn0}$ and $K_{0ll0}$
through
\begin{eqnarray}
\nonumber
|\alpha^{ce}(0)|^2&=&\fmn{1}{4}\left[1+2K_{0nn0}(\pi)+K_{0ll0}(\pi)\right]
\left(\frac{\dd\sigma}{\dd t}\right)_{\!np\to np},\\
\nonumber
|\beta^{ce}(0)|^2&=&\fmn{1}{4}\left[1-K_{0ll0}(\pi)\right]
\left(\frac{\dd\sigma}{\dd t}\right)_{\!np\to np},\\
|\epsilon^{ce}(0)|^2&=&\fmn{1}{4}\left[1-2K_{0nn0}(\pi)+K_{0ll0}(\pi)\right]
\left(\frac{\dd\sigma}{\dd t}\right)_{\!np\to np}\!\!.
\label{observables1}
\end{eqnarray}
It should be noted that the corresponding results for the
non-charge-exchange amplitudes, \emph{i.e.}\ without the
interchange operator $\mathcal{P}_{p\leftrightarrow n}$ in
Eq.~(\ref{Mce}), have a similar structure but with the $K$ being
replaced by the depolarisation parameters $D$. This arises because
of the different assignment of the labels \emph{scattered} and
\emph{recoil} to the final particles in the two decompositions. Of
course, the \emph{elastic} and \emph{charge-exchange}
representations must lead to the same physics for elastic
neutron-proton scattering, with merely interpretational
differences.

Using the above relations in association with the phase shift
predictions from the SAID database~\cite{SAID}, one sees that the
spin-independent term $|\alpha^{ce}(0)|^2$ should contribute less
than 10\% to the forward charge-exchange cross section between say
200\,MeV up to the limit of the SAID analysis at 1.3\,GeV. In
contrast, there is relatively little spin flip between the initial
and final neutrons, \textit{i.e.}\ the $|a(\pi)+b(\pi)|^2$ term
dominates. One has therefore to be very careful to specify clearly
the meaning of any statement comparing the magnitudes of the
spin-flip and spin-independent contributions in backward
neutron-proton elastic scattering.
%
%
\section{Charge exchange on the deuteron}
\label{deuteron} \setcounter{equation}{0}

\begin{figure}[htb]
\begin{center}
\centerline{\epsfxsize=6cm{\epsfbox{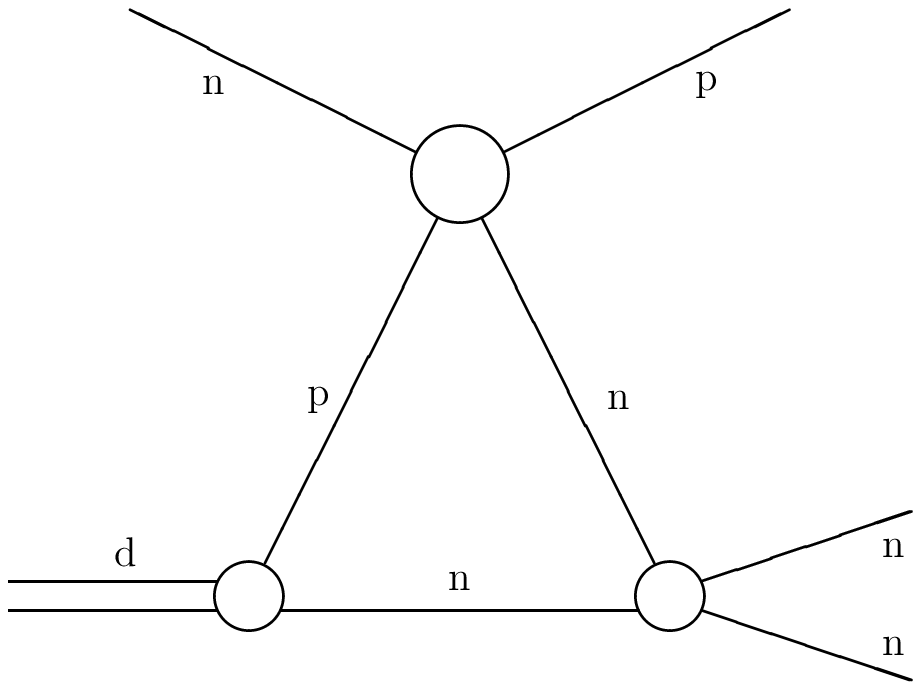}}}
\end{center}
\vspace{-0.2cm} \caption{\label{diagram} Impulse approximation
diagram for nucleon charge exchange on the deuteron.}
\end{figure}

In single-scattering (impulse) approximation, the charge exchange
$nd\to p\,\{nn\}$ reaction on the deuteron is thought of as a
$np\to pn$ reaction with a spectator neutron\footnote{Since the
deuteron has isospin-zero, the description of the $pd\to
n\,\{pp\}$ reaction is formally identical.}. Initially the
neutron-proton pair is bound in the deuteron and the two emerging
neutrons are subject to a final state interaction, as illustrated
diagramatically in Fig.~\ref{diagram}. If $\vec{k}$, the relative
momentum in the $nn$ system, and hence the excitation energy
$E_{nn}=k^2/m$, are small, the final neutrons are in the $^{1\!}S_0$
state. The reaction therefore acts as a spin-isospin filter going
from the $(^{3\!}S_1,^{3\!\!}D_1)$ of the deuteron to the final
$^{1\!}S_0$ of the dineutron. Furthermore, at low momentum
transfers $\vec{q}=\vec{k}_f-\vec{k}_i$ between the initial
neutron and final proton other final states are only weakly
excited. Under such conditions the $nd\to p\,\{nn\}$ differential
cross section depends but weakly upon the spin-independent
amplitude $\alpha^{ce}$ of Eq.~(\ref{Mce}).

The above features were put on a simple quantitative basis through
the use of a sum rule by Dean~\cite{Dean} and this was extended to
polarisation observables by Bugg and Wilkin~\cite{BW}. The matrix
element of the transition is of the form~\cite{Carbonell}
\begin{eqnarray}
\label{impulse}%
\lefteqn{\mathcal{F}(\vec{k}_f,\vec{k}_i;S,\nu_f,M,m_p,m_n) =}\\
\nonumber
&&\hspace{-4mm}\langle\Psi_{nn,\vec{k}}^{(-)};S,\nu_f,m_p|M^{ce}
(\vec{k}_f,\vec{k}_i)
\exp(\half i\vec{q}\cdot\vec{r})|\Phi_d^M,m_n\rangle\,,
\end{eqnarray}
where $\Phi_d(\vec{r})$ is the deuteron wave function in
configuration space and $\Psi_{nn,\vec{k}}^{(-)}(\vec{r})$ the
corresponding for the low energy $nn$ final state of spin $S$ and
projection $\nu_f$. The magnetic quantum numbers of the initial
neutron and deuteron and the final proton are denoted by $m_n$,
$M$, and $m_p$, respectively.

The unpolarised cross section is normalised such that
\begin{equation}
\label{normd} \frac{\dd4\sigma}{\dd{t}\,\dd^3k}=\fmn{1}{6}\,
\textrm{Tr}\left\{\mathcal{F}^{\dagger}\mathcal{F}\right\},
\end{equation}
where the trace is over all the spin projections.

Dean noted that, if one summed Eq.~(\ref{normd}) over all
excitation energies of the dineutron, one could obtain a sum rule
that did not depend upon the details of the low energy $nn$
interaction. This is a high energy approximation and, for it to be
valid, the available phase space must be so large as not to
disturb the convergence of the sum rule, as discussed below.

Specialising to the case of interest here, \emph{viz.},
$\theta_{np}=0^{\circ}$, $q\approx 0$, the sum rule for the
differential cross section reduces to
\begin{eqnarray}
\nonumber \lefteqn{\frac{\dd\sigma}{\dd{t}}(nd\to p\{nn\})=\int
\frac{\dd^4\sigma}{\dd{t}}\,\dd^3k}\\ &&=
\fmn{2}{3}\left(2|\beta^{ce}(0)|^2+|\epsilon^{ce}(0)|^2\right).
\end{eqnarray}

For completeness, we give also the corresponding sum rule for the
deuteron tensor analysing power $t_{20}$;
\begin{eqnarray}
\nonumber \lefteqn{t_{20}\,\frac{\dd\sigma}{\dd{t}}(nd\to
p\{nn\})}\\&&= \frac{2\sqrt{2}}{3}\,\left(1-\fmn{9}{10}P_D\right)
\left(|\beta^{ce}(0)|^2-|\epsilon^{ce}(0)|^2\right),
\label{tensorrule}
\end{eqnarray}
where $P_D$ is the deuteron $D$-state probability.

The Dean sum rule may then be written in terms of $np\to np$
elastic scattering observables as
\begin{eqnarray}
\nonumber R_{np}(0)&=&\left.\frac{\dd\sigma(nd\to
p\{nn\})/\dd{t}}{\dd\sigma(np\to pn)/\dd{t}}\right|_{q=0}\\
\nonumber&&\hspace{-1cm}= \frac{2}{3}
\frac{2|\beta^{ce}(0)|^2+|\epsilon^{ce}(0)|^2}
{|\alpha^{ce}(0)|^2+2|\beta^{ce}(0)|^2+|\epsilon^{ce}(0)|^2}\\
&&\hspace{-1cm}=
\fmn{1}{6}\left[3-K_{0ll0}(\pi)-2K_{0nn0}(\pi)\right].
\label{Rnptheory}
\end{eqnarray}

\begin{figure}[htb]
\begin{center}
\centerline{\epsfxsize=8cm{\epsfbox{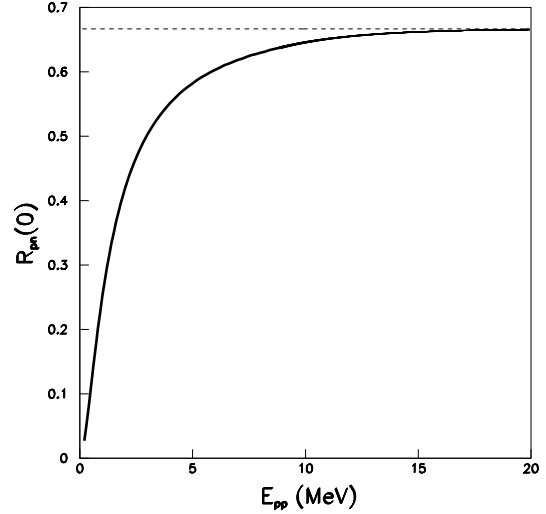}}}
\end{center}
\vspace{-0.2cm} \caption{Convergence of the sum rule for the
$pd\to n\{pp\}$ reaction at $q=0$ as a function of the excitation
energy in the final $pp$ system. It has been assumed that the
spin-non-flip amplitude $\alpha^{ce}=0$ to make the limiting value
$\frac{2}{3}$, though the rate of convergence is similar in all
cases. The evaluation has been carried out using the de Toureill
and Sprung potential~\cite{Sprung} to describe the low energy
nucleon-nucleon systems, as explained in Ref.~\cite{Carbonell}.
\label{sumrulebye}}
\end{figure}

The sum rule of Eq.~(\ref{Rnptheory}) is very effective at medium
and high energies because it converges so quickly. This is
illustrated in Fig.~\ref{sumrulebye}, where the impulse
approximation cross section of Eq.~(\ref{normd}) has been
evaluated for the $pd\to n\{pp\}$ case~\cite{Carbonell} and
integrated numerically over the $pp$ excitation energy. Although a
specific nucleon-nucleon potential was used~\cite{Sprung}, the
rate of convergence depends little upon this choice and less than
1\% of the sum rule remains beyond $E_{pp}\approx 15$\,MeV. The
rate of convergence under the specific conditions of the Dubna
$d(n,p)nn$ experiment~\cite{Sharov08} has also been
discussed by Ladygina~\cite{Ladygina}.

Deviations might be expected from the sum rule at lower energies
where the phase space available does not allow an unimpeded
integration over $E_{nn}$. As discussed in section~\ref{data},
this has only a marginal effect on the saturation of the sum rule
for incident energies above 50\,MeV. More critical, though, is the
fact that at low energies experimentalists generally put a more
severe cut on the momentum of the final proton in the $nd\to
p\{nn\}$ reaction to minimise the contributions from diagrams
other than those of the impulse approximation and this reduces the
value obtained for $R_{np}(0)$. Although we later report results
at low energies, there is no reason to believe that the impulse
approximation of Fig.~\ref{diagram} should dominate there and a
full three-body calculation has to be undertaken to interpret
these results.

The most significant correction to the sum rule at high energies
comes from the shadow effect~\cite{Glauber}, which will typically
reduce the value of $R_{np}(0)$ by about 5\%.
%
%
\section{Experimental data and theoretical comparison}
\label{data} \setcounter{equation}{0}

\begin{table}[hbt]
\begin{center}
\caption{Summary of the available experimental data on the
$R_{np}(0)$ ratio measured using the $nd\to p\,\{nn\}$ reaction.
The error bars reflect both the statistical and systematic
uncertainties.\label{table1}}
\begin{tabular}{|r|c|c|r|}
\hline
&&&\\
$T_{\rm kin}$~~ & $R_{np}(0)$ & Year & Ref.\\
 (MeV)    &               &    &      \\
\hline
  13.9 & $0.185~~~~~~~~~~~$   & 1965 &\cite{VOI65}  \\
  90.0 & $0.397 \pm 0.044$    & 1951 &\cite{POW51}  \\
 152.0 & $0.650 \pm 0.100$    & 1966 &\cite{MEA66}  \\
 200.0 & $0.553 \pm 0.030$    & 1962 &\cite{DZH62}  \\
 270.0 & $0.710 \pm 0.021$    & 1952 &\cite{CLA52}  \\
 299.7 & $0.652 \pm 0.033$    & 1988 &\cite{PAG88}  \\
 319.8 & $0.643 \pm 0.032$    & 1988 &\cite{PAG88}  \\
 339.7 & $0.637 \pm 0.032$    & 1988 &\cite{PAG88}  \\
 359.6 & $0.626 \pm 0.031$    & 1988 &\cite{PAG88}  \\
 379.6 & $0.641 \pm 0.032$    & 1988 &\cite{PAG88}  \\
 380.0 & $0.200 \pm 0.035$    & 1955 &\cite{DZH55}  \\
 399.7 & $0.610 \pm 0.031$    & 1988 &\cite{PAG88}  \\
 419.8 & $0.623 \pm 0.031$    & 1988 &\cite{PAG88}  \\
 440.0 & $0.630 \pm 0.032$    & 1988 &\cite{PAG88}  \\
 460.1 & $0.611 \pm 0.031$    & 1988 &\cite{PAG88}  \\
 480.4 & $0.608 \pm 0.030$    & 1988 &\cite{PAG88}  \\
 500.9 & $0.592 \pm 0.030$    & 1988 &\cite{PAG88}  \\
 521.1 & $0.604 \pm 0.030$    & 1988 &\cite{PAG88}  \\
 539.4 & $0.617 \pm 0.031$    & 1988 &\cite{PAG88}  \\
 550.0 & $0.589 \pm 0.046$    & 2007 &\cite{Sharov08}  \\
 557.4 & $0.632 \pm 0.032$    & 1988 &\cite{PAG88}  \\
 710.0 & $0.483 \pm 0.080$    & 1960 &\cite{LAR60}  \\
 794.0 & $0.560 \pm 0.040$    & 1978 & \cite{BON78} \\
 800.0 & $0.554 \pm 0.023$    & 2007 & \cite{Sharov08} \\
1000\phantom{.0} & $0.553 \pm 0.026$     & 2007 & \cite{Sharov08} \\
1200\phantom{.0} & $0.551 \pm 0.022$     & 2007 & \cite{Sharov08} \\
1400\phantom{.0} & $0.576 \pm 0.038$     & 2007 & \cite{Sharov08} \\
1800\phantom{.0} & $0.568 \pm 0.033$     & 2007 & \cite{Sharov08} \\
2000\phantom{.0} & $0.564 \pm 0.045$     & 2007 & \cite{Sharov08} \\
\hline

\end{tabular}
\end{center}
\end{table}

%
%
\begin{table}[hbt]
\begin{center}
\caption{Summary of the available experimental data on the
$R_{np}(0)$ ratio measured using the $pd\to n\,\{pp\}$ reaction.
The error bars reflect both the statistical and systematic
uncertainties. \label{table2}}
\end{center}
\begin{center}
\begin{tabular}{|r|c|c|c|}
\hline
&&&\\
$T_{\rm kin}$~~ & $R_{np}(0)$ & Year & Ref.\\
 (MeV)    &               &      &       \\
\hline
 ~13.5 & $0.180~~~~~~~~~~~$ & 1959 &\cite{WON59}  \\
 ~30.1 & $0.141 \pm 0.035$ & 1967 &\cite{BAT66}  \\
 ~50.0 & $0.240 \pm 0.060$ & 1967 &\cite{BAT66}  \\
 ~95.0 & $0.480 \pm 0.030$ & 1953 &\cite{HOF53}  \\
 ~95.7 & $0.587 \pm 0.029$ & 1967 &\cite{LAN67}  \\
 135.0 & $0.652 \pm 0.154$ & 1965 &\cite{EST65}  \\
 143.9 & $0.601 \pm 0.057$ & 1967 &\cite{LAN67}  \\
 647.0 & $0.600 \pm 0.080$ & 1976 &\cite{BJO76} \\
 800.0 & $0.660 \pm 0.080$ & 1976 & \cite{BJO76} \\
\hline
\end{tabular}
\end{center}
\end{table}

\begin{table}[hbt]
\begin{center}
\caption{Summary of the available experimental data on the
$R_{np}(0)$ ratio measured using the $dp\to \{pp\}\,n$ reaction.
The kinetic energy quoted here is the energy per nucleon. The
error bars reflect both the statistical and systematic
uncertainties. \label{table3}}
\begin{tabular}{|c|c|c|c|}
\hline
&&& \\
$T_{\rm kin}$ & $R_{np}(0)$ & Year & Ref.\\
 (MeV)    &              &      &       \\
\hline
 977.0 & $0.430 \pm 0.220$ & 1975 & \cite{ALA75}  \\
 977.0 & $0.650 \pm 0.120$ & 2002 & \cite{GLA02} \\
\hline
\end{tabular}
\end{center}
\end{table}
%
%

The cross section ratio $R_{np}(0)$ can be investigated using
either the $nd \to p\{nn\}$, $pd\to n\{pp\}$, or the $dp\to \{pp\}
n$ reaction and the experimental results in the three cases are
summarised, respectively, in Tables~\ref{table1}, \ref{table2},
and \ref{table3}. In many cases the results had to be read from
graphs and this is especially true of some of the very old data.
However, because of their age, such data tend to be at the lower
energies, which are of less interest for the amplitude
reconstruction.

There are more measurements of the $nd \to p\{nn\}$ reaction than
of $pd\to n\{pp\}$ and the error bars are also generally smaller.
This reflects the relative difficulty in the two types of
experiment. If a good quality neutron beam is available, the
measurement of the relative strengths of the proton spectra from
deuterium and hydrogen targets is comparatively
\textit{straightforward}. The alternative of measuring the neutron
in the final state presents far more difficulties. The third
technique of using a deuteron beam has only been attempted once
because in this case one has to measure both final fast protons
over a large range of phase space. This can be done using a bubble
chamber and the value~\cite{GLA02} quoted in Table~\ref{table3}
represents the result of increased statistics and a different
analysis compared to that of the previous
publication~\cite{ALA75}. A particular criticism here is that the
free neutron-proton data have to be taken from an entirely
different source so that there can be no cancellations between any
systematic errors.

Although, as shown by Fig.~\ref{sumrulebye}, the sum rule
converges quite fast, there is nevertheless a problem at low
energies as to where to cut the tail in the proton spectrum and, in
certain cases, it is likely that the choice has not allowed for a
full saturation of the sum rule.

\begin{figure}[hbt]
\centering
\includegraphics[width=0.9\columnwidth,clip]{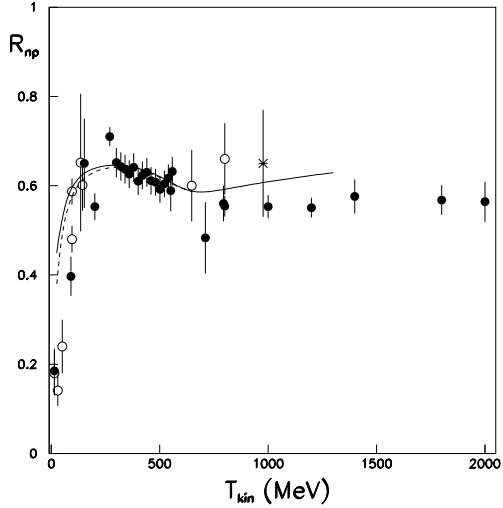}
\caption{Experimental data on the $R_{np}(0)$ ratio at zero
momentum transfer. The closed circles are from the $(n,p)$ data of
Table~\ref{table1}, the open circles from the $(p,n)$ data of
Table~\ref{table2}, and the cross from the $(d,2p)$ datum of
Table~\ref{table3}. These results are compared to the Dean
sum-rule predictions of Eq.~(\ref{Rnptheory}) using the current
SAID solution, which is available up to a laboratory kinetic
energy of 1.3\,GeV. The dashed curve takes into account the
limited phase space available at the lower
energies.\label{sumrule}}
\end{figure}

The different measurements of $R_{np}(0)$ are reported in
Fig.~\ref{sumrule} with one controversial point~\cite {DZH55}
being omitted. Also shown are the estimates of
Eq.~(\ref{Rnptheory}), evaluated using the current SAID
solution~\cite{SAID}. In these predictions, no account has been
taken of the shadow effect~\cite{Glauber}, which would reduce the
limiting value to typically 0.63/0.64. The agreement with the
phase shift predictions is generally satisfactory, though there
are suggestions from the new data~\cite{Sharov08} that there might
might be some underestimate of the relative strength of the
$|\alpha^{ce}(0)|^2$ contribution at the higher energies.

At low energies the predictions fall steeply. In part this is
related to the behaviour of the $np$ amplitudes but one has also
to take account of the fact that the smaller available phase space
does not allow the sum rule to be fully saturated. However, even
when this effect is included, by integrating numerically the
impulse approximation, the effect is comparatively small and does
not account for the much steeper drop in the experimental data. Of
much more importance there is the experimental cut in the recoil
proton momentum and, of course, the deviations from the naive
impulse approximation at low energies.

%
%
\section{Summary and conclusions}
\label{summary} \setcounter{equation}{0}

The ratio $R_{np}(0)$ of the forward differential cross section
for charge exchange on the deuteron to that on the nucleon is a
very robust observable since many uncertainties drop out between
the two measurements. It is therefore not surprising that one gets
a consistent picture of the energy dependence of this quantity
from measurements of the $nd \to p\{nn\}$, $pd\to n\{pp\}$, and
$dp\to \{pp\} n$ reactions. Furthermore, we have shown that, from
about 300\,MeV up to 1.3\,GeV, where the $np$ phase shift analyses
terminate, the predictions for $R_{np}(0)$ largely agree with the
available experimental data. Since measurements of $R_{np}(0)$
have now been carried out up to 2\,GeV~\cite{Sharov08}, it is
appropriate to consider whether a neutron-proton elastic amplitude
reconstruction could be performed using these results.

In the backward direction, we see from Eq.~(\ref{identification2})
that there are only three (complex) amplitudes, $\alpha^{ce}(0)$,
$\beta^{ce}(0)$, and $\epsilon^{ce}(0)$. In principle, as shown by
Eq.~(\ref{sigtot2}), the imaginary parts of these amplitudes can
be fixed by measurements of the spin dependence of the $pp$ and
$np$ total cross sections. According to the available phase shift
analyses, these quantities are much smaller than the real parts
and so the error bars will be relatively large, even if all the
measurements were available. The values of $|\alpha^{ce}(0)|^2$
and $2|\beta^{ce}(0)|^2+|\epsilon^{ce}(0)|^2$ would be fixed by
the combined measurement of $R_{np}(0)$ and the free $np\to np$
differential cross section.

Measurements have been carried out on the $t_{20}$ tensor
analysing power of the
$d\hspace{-3.2mm}\stackrel{\to}{\phantom{p}}\hspace{-1.4mm}p\to
\{pp\}n$ reaction~\cite{ELL87,KOX93,CHI06}. Rather than using the
sum rule of Eq.~(\ref{tensorrule}), these data were taken with an
excitation energy in the $pp$ system so low that the final
$^{1\!}S_0$ system dominates or where one could correct from
contamination for the $pp$ $P$-waves. Under such conditions,
\begin{equation}
t_{20}=\sqrt{2}\left(\frac{|\beta^{ce}(0)|^2-|\epsilon^{ce}(0)|^2}
{2|\beta^{ce}(0)|^2+|\epsilon^{ce}(0)|^2}\right)\cdot
\end{equation}
Furthermore, a measurement of the transverse spin correlation with
an incident vector polarised deuteron on a polarised hydrogen
target would give~\cite{Barbaro}
\begin{equation}
C_{n,n}(0)=\frac{-2\textit{Re}(\beta^{ce}(0)^*\epsilon^{ce}(0))}
{2|\beta^{ce}(0)|^2+|\epsilon^{ce}(0)|^2}\cdot
\end{equation}

It is therefore clear that, even with very good data on these
seven parameters, there would still be two discrete ambiguities,
\textit{viz.}\ the signs of the $\textit{Re}(\alpha^{ce}(0)$ and
$\textit{Re}(\beta^{ce}(0)$. However, since there are extensive
data on the unpolarised cross section difference of
Eq.~(\ref{sigtot2}), the application of forward dispersion
relations will certainly be sufficient to determine at least the
sign of $\textit{Re}(\alpha^{ce}(0)$. Regarding the other
ambiguity, if $\textit{Re}(\beta^{ce}(0)$ changes sign then this
would be reflected in the value of $t_{20}(0)$, which has been
found to have the same sign throughout the measured
range~\cite{ELL87,KOX93,CHI06}. In principle, therefore, it seems
that a direct amplitude construction of $np\to np$ in the backward
direction is feasible without having to measure differential cross
sections with polarised protons and neutrons, although these would
clearly enhance the precision of any analysis.

For obvious experimental reasons, the measurements of deuteron
charge exchange leading to the low excitation energy $pp$ system
have so far been carried out with a polarised deuteron
beam~\cite{ELL87,KOX93,CHI06}. This lowers the limit on the energy
per nucleon. However, a polarised hydrogen/deuterium gas target is
now available at the ANKE facility of the COSY-J\"ulich storage
ring~\cite{SPIN}. Using solid state telescopes placed within the
target chamber, the low energy protons from the
$p\hspace{-3mm}\stackrel{\to}{\phantom{p}}\hspace{-1.2mm}d\hspace{-2.7mm}\stackrel{\to}{\phantom{p}}\to n\{pp\}$
reaction can be measured with high precision. This will
allow the values of $t_{20}(0)$ and $C_{n,n}(0)$ to be
investigated up to the maximum proton beam energy of 2.9\,GeV.

%
%

\begin{acknowledgement}
This work has resulted from an active and productive collaboration
with the Dubna DELTA-SIGMA group of Ref.~\cite{Sharov08}, for
which we are very appreciative. We are also grateful to J.~Ludwig
for furnishing us with a copy of Ref.~\cite{PAG88}. One of the
authors (CW) wishes to thank the Czech Technical University, where
this work was initiated, for hospitality and support.
\end{acknowledgement}
%
%

\end{document}